\documentclass[preprint]{vgtc}               % preprint
\graphicspath{{figures/}{pictures/}{images/}{./}} % where to search for the images

\usepackage{times}                     % we use Times as the main font
\usepackage{tabu}                      % only used for the table example
\usepackage{booktabs}                  % only used for the table example
\usepackage{lipsum}                    % used to generate placeholder text
\usepackage{mwe}                       % used to generate placeholder figures

\usepackage{mathptmx}                  % use matching math font

\onlineid{0}

\vgtccategory{Research}

\vgtcinsertpkg

\preprinttext{To appear as a short paper in the upcoming IEEE VIS 2024 conference.}

\title{Demystifying Spatial Dependence: Interactive Visualizations for Interpreting Local Spatial Autocorrelation}

\author{Lee Mason\thanks{e-mail: masonlk@nih.gov}\\ %
       \parbox{1.4in}{\scriptsize \centering National Cancer Institute \\ Queen's University Belfast} %
\and Blánaid Hicks\thanks{e-mail: b.hicks@qub.ac.uk}\\ %
     \scriptsize Queen's University Belfast %
\and  Jonas Almeida\thanks{e-mail: jonas.dealmeida@nih.gov}\\ %
      \scriptsize National Cancer Institute}

\abstract{
   The Local Moran's I statistic is a valuable tool for identifying localized patterns of spatial autocorrelation. Understanding these patterns is crucial in spatial analysis, but interpreting the statistic can be difficult. To simplify this process, we introduce three novel visualizations that enhance the interpretation of Local Moran's I results. These visualizations can be interactively linked to one another, and to established visualizations, to offer a more holistic exploration of the results. We provide a JavaScript library with implementations of these new visual elements, along with a web dashboard that demonstrates their integrated use. 
} % end of abstract

\keywords{Spatial, spatial clustering, spatial autocorrelation, geospatial, GIS, interactive visualization, visual analytics, Moran's I, local indicators of spatial association} 

\begin{document}

\firstsection{Introduction}

\maketitle

``Everything is related to everything else, but near things are more related than distant things''. This is Tobler’s first law of geography, a defining principle of spatial analysis \cite{toblerComputerMovieSimulating1970,millerToblerFirstLaw2004}. One of the most popular analytical concepts in spatial analysis, and perhaps the one that best encapsulates Tobler’s law, is spatial autocorrelation \cite{getisSpatialAutocorrelation2009, anselinLocalIndicatorsSpatial1995}. Spatial autocorrelation is a measure of the extent to which spatially-referenced values correlate between adjacent locations. 
Spatial autocorrelation is challenging to evaluate from visual representations of the data alone \cite{beechamMapLineUpsEffects2017a, klippelInterpretingSpatialPatterns2011}, and thus there are several statistical measures which formalize the concept numerically. The most popular of these is Moran's I, a basic quantification which contrasts spatial autocorrelation to spatial randomness \cite{moranNotesContinuousStochastic1950}. It was originally conceived as a global property of the dataset, quantifying the presence of spatial autocorrelation but not indicating which points in the dataset are most and least correlated with their neighbours. The Local Moran’s I statistic was introduced to address the need for a granular assessment by breaking down the global spatial autocorrelation value into a separate value for each observation \cite{anselinLocalIndicatorsSpatial1995}. The Local Moran's I statistic can be used to detect spatial clusters and spatial outliers, two important concepts in spatial analysis. Both Moran’s I and Local Moran’s I are popular and versatile statistics which have been used for a variety of tasks across disciplines, including evaluating the role of gender in the spatial distribution of obesity rates \cite{schuurmanAreObesityPhysical2009}, tracking the spread of COVID-19 \cite{ghoshSpatioTemporalAnalysis2020}, studying neighbourhood deprivation at the county-level \cite{andrewsGeospatialAnalysisNeighborhood2020}, and more \cite{zhangUseLocalMoran2008, yuanUsingLocalMoran2018, goovaertsDetectionTemporalChanges2005, boneGISbasedRiskRating2013, xieDetectingTrafficAccident2013}.

Interpreting the Local Moran's I statistic can be challenging due to the complex factors involved in its calculation and subsequent analysis. This is especially problematic in exploratory environments, where a rapid understanding of the results is imperative \cite{cuiVisualAnalyticsComprehensive2019a}. Visualization can help overcome this challenge by presenting complex results in a manner which is more suitable for rapid interpretation \cite{anselinMoranScatterplotESDA2019}. While there is an extensive landscape of spatial autocorrelation plot designs, there are notable gaps in the current body of work. For instance, there is no single plot that represents all of the elements involved in a Local Moran's I calculation.

In this work, we introduce three novel visual designs which extend the landscape of local spatial autocorrelation visualization, each focusing on a different and complementary aspect of the Local Moran's I results.
We also suggest a way to link these plots through the use of shared axes, proximate overlays, and user interaction. 
We provide an implementation of the novel visual elements in an open-source JavaScript library and complementary web dashboard, both of which run entirely in the user's web-browser. These are available at \url{https://episphere.github.io/moranplot}.

%%%%%%%%%%%%%%%%%%%%%%%%%%%%%%%%%%%%%%%%%%%%%%%%%%%%%%%%%%%%%%%%
%%%%%%%%%%%%%%%%%%%%%%%%%%%%%%%%%%%%%%%%%%%%%%%%%%%%%%%%%%%%%%%%
\section{Local Moran's I}

The Local Moran's I statistic is a measure of local spatial autocorrelation, numerically encapsulating how each location in a geospatial dataset correlates with its immediate neighbors. Calculating Local Moran's I requires a dataset of attribute values $X = [x_1, ..., x_n]$ and a $n \times n$ weight matrix $W$ which defines the spatial relationship between locations. For instance, in this work's example dataset, each attribute value represents an age-adjusted cancer mortality rate for a US county, and the weight matrix indicates whether a pair of counties share a border (1 if they do, 0 if they don't). For simplicity of calculation, the attribute values are typically z-score normalized (giving $Z$) and the weight matrix is typically row-normalized (giving $W'_{i,j}$). With that, the Local Moran's I for a dataset is calculated as:
%%%
\begin{equation} \label{eq_lisa}
I_i = \frac{z_i \cdot lag_i}{n-1} = \frac{z_i \cdot \sum_j^n W'_{ij} \cdot z_j}{n-1}
\end{equation}
Where $z_i$ is the z-score normalized value for location $i$ and $W'_{i,j}$ is the quantified spatial relationship between location $i$ and location $j$. Note the spatial lag term $lag_i = \sum_j^n W'_{ij} \cdot z_j$, this is a useful concept in spatial analysis and plays an important role in Local Moran's I interpretation and visualization. 

Interpreting a Local Moran's I value typically involves determining its significance and assigning an appropriate label. To determine significance for a result, the dataset is permuted a number of times: on each permutation the focal location's attribute value remains fixed while the other values are shuffled. The statistic is calculated for each permutation, creating an empirical distribution which the observed statistic can be compared against. A pseudo p-value is calculated from the number of permuted statistic values which are more extreme than the observed value. On it's own, this only shows whether a location exhibits positive or negative spatial autocorrelation. To detect spatial clusters and outliers, it is necessary to further refine the results using their z-score normalized attribute value and their lag value. Positive spatial autocorrelation values (spatial clusters) are labeled as ``high-high'' (positive value and lag) or ``low-low'' (negative value and lag). Negative spatial autocorrelation (spatial outliers) are labeled as ``high-low'' (positive value, negative lag) or ``low-high'' (negative value, positive lag).

%%%%%%%%%%%%%%%%%%%%%%%%%%%%%%%%%%%%%%%%%%%%%%%%%%%%%%%%%%%%%%%%
%%%%%%%%%%%%%%%%%%%%%%%%%%%%%%%%%%%%%%%%%%%%%%%%%%%%%%%%%%%%%%%%
\section{Local Moran's I Visualization Elements} \label{sec_elements}

To better understand the design requirements of an effective Local Moran's I visualization, we have identified a number of core elements involved in the calculation and interpretation of the statistic. The statistic's formula directly provides three of these elements: the attribute value at the focal location (\textbf{E1}), the attribute values at neighboring locations (\textbf{E2}), and the neighbor weights from the weight matrix (\textbf{E3}). The spatial lag (\textbf{E4}) is a combination of E2 and E3 but because it is an important spatial concept it is worth considering in its own right. The Local Moran's I statistic value (\textbf{E5}) is the product of E1 and E4; the sign indicating positive or negative spatial autocorrelation. The magnitude of the statistic must be interpreted in relation to a reference distribution (\textbf{E6}), which is typically calculated empirically. A pseudo p-value (\textbf{E7}) can be generated by referencing E5 against E6. 
Finally, the threshold values for significance (\textbf{E8}) are a concrete way to interpret the statistic's meaning and label the result.

%%%%%%%%%%%%%%%%%%%%%%%%%%%%%%%%%%%%%%%%%%%%%%%%%%%%%%%%%%%%%%%%
%%%%%%%%%%%%%%%%%%%%%%%%%%%%%%%%%%%%%%%%%%%%%%%%%%%%%%%%%%%%%%%%
\section{Related Work}

Visualization has been integral to Moran's I and Local Moran's I analysis since its inception, with numerous plot designs proposed over the years \cite{anselinLocalIndicatorsSpatial1995, anselinMoranScatterplot1996, anselinInteractiveTechniquesESDA1996, vasilievVisualizationSpatialDependence2020, wartenbergMultivariateSpatialCorrelation1985}. However, significant research gaps remain. Notably, no existing plot design comprehensively represents all the key elements outlined in \cref{sec_elements} --- most focus on only a small subset. For example, the influential Moran scatterplot represents only E1 and E4, with each local result plotted as a point whose x and y coordinates correspond to the location's z-normalized attribute value and spatial lag, respectively \cite{anselinMoranScatterplot1996}. The Moran drop plot extends this by vertically connecting each significant point in the Moran's scatterplot to its nearest critical value (a variant on E8) \cite{westerholtExtendingMoranScatterplot2024}. Part of the reason for the gap here is that both plots, along with other variants of the Moran scatterplot, primarily focus on conveying global structure rather than detailed information about local results. Few plot designs prioritize a single result. Two exceptions are spatial lag bar charts and pie charts, both of which display the ratio of a location's attribute value to its spatial lag \cite{anselinInteractiveTechniquesESDA1996}. However, this is only two core elements, and the ratio itself isn't directly involved in the Local Moran's I calculation, limiting the value of these plots for interpreting results. 

Another feature typically neglected from Moran's I visualization is a convenient means to inspect the results in relation to their corresponding geospatial locations. One exception to this is the Moran location scatterplot, a variant of the Moran scatterplot which maintains the plot's quadrants but
within those quadrants points are positioned according to their geographic location instead of their attribute value and spatial lag \cite{negreirosComprehensiveFrameworkExploratory2010}. While this represents geospatial information, it does so by sacrificing an explicit representation of E1 and E4. Similarly, no existing plot design explicitly connects neighboring results to each other, which is a natural and useful way to engage with the spatial clusters returned from a Local Moran's I analysis.

%%%%%%%%%%%%%%%%%%%%%%%%%%%%%%%%%%%%%%%%%%%%%%%%%%%%%%%%%%%%%%%%
%%%%%%%%%%%%%%%%%%%%%%%%%%%%%%%%%%%%%%%%%%%%%%%%%%%%%%%%%%%%%%%%
\section{Moran Plots}

To assist with the interpretation of Local Moran's I results, we have designed three novel plots which convey different elements of the results. These plots were designed from our own observations of gaps in the visualization landscape, and refined using feedback from novice and expert target users. We have provided these plots in a JavaScript library which is documented at \url{https://github.com/episphere/moranplot}.

\subsection{Moran Dual-Density Plot}

\begin{figure}[bt]
  \centering 
  \includegraphics[width=.99\columnwidth, alt={A figure showing the Moran dual-density plot}]{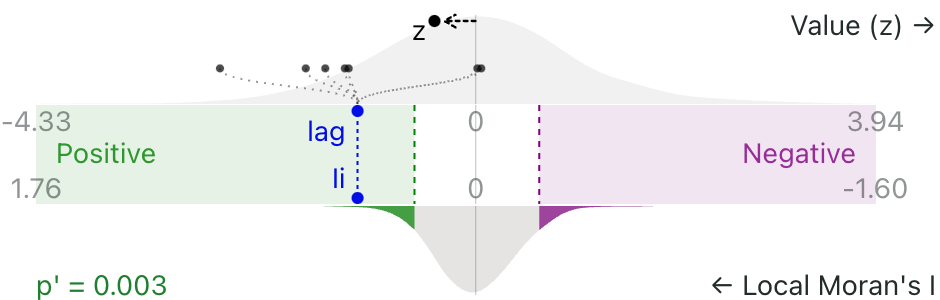}
  \caption{%
    The Moran dual-density plot showing a negative z-score normalized attribute value and a negative spatial lag. The blue points are within the positive significance area, indicating that the statistic exhibits significant positive spatial autocorrelation.
  }
  \label{fig:dual_density}
\end{figure}

The Moran dual-density plot was designed to visually consolidate the key components involved in calculating and interpreting Local Moran's I values (see \cref{fig:dual_density}). It features two connected density plots. The first (on top) displays the distribution of z-score normalized attribute values within the spatial dataset, serving as a reference for the focal location's attribute value (E1), which is represented as a point on the top axis. This point is connected to the $z = 0$ value by a dashed line, which emphasizes the value's most important characteristic: it's sign. Somewhat counter-intuitively, the magnitude of this value has no influence on the significance or interpretation of the statistic. We include this information in the plot regardless for the sake of completeness and to maintain a consistent means of representing attribute values. Additionally, representing the magnitude visually but not linking it to other plot glyphs reinforces the counter-intuitive lack of role the attribute value magnitude plays in the result interpretation. 
%%%
Also present on the top part of the plot are the z-score normalized attribute values for each neighbor (E2), which are represented as points with radius proportional to the corresponding weight in the weight matrix (E3). The spatial lag (E4) is represented as a colored point on the x-axis. Each of the neighbor points are connected to this point with dashed lines, illustrating their role in the calculation of the spatial lag. 
%%%
The point representing the spatial lag (E4) is connected to a point representing the Local Moran's I value (E5), demonstrating the linear relationship between the two elements. A density plot on the bottom axis visualizes the distribution of permuted Local Moran's I values. Significance thresholds for positive and negative local spatial autocorrelation (E8) are marked with colored areas derived from the permuted values. The pseudo p-value (E7) is shown on the bottom left. The relative orientation of these areas can change based on the sign of the normalized attribute value, and therefore we have color coded and labelled each area to avoid confusion. There are two possible color modes, one showing negative and positive spatial autocorrelation, and another representing the label assignment. The former mode (see \cref{fig:dual_density}) is closer to the direct interpretation of the statistic, whereas the latter (see \cref{fig:dash}) is closer to how the statistic is used in practice (for spatial cluster and outlier detection). 

We chose to base this design around two horizontally connected axes to emphasize the linear relationship between the spatial lag and the Local Moran's I statistic, a relationship essential to the statistic's interpretation. The accompanying density plots contextualize these values to aid in understanding of their distribution and significance. The plot's distinct segments allow for an appropriate separation of information. We chose to jointly represent the spatial lag and statistic value in the central segment to explicitly convey their close relationship, which also allows the significance cut-offs to be presented in direct relation to both elements.

\subsection{Moran Network Scatterplot}

\begin{figure}[h]
  \centering 
  \includegraphics[width=.61\columnwidth, alt={A figure showing the Moran network scatterplot}]{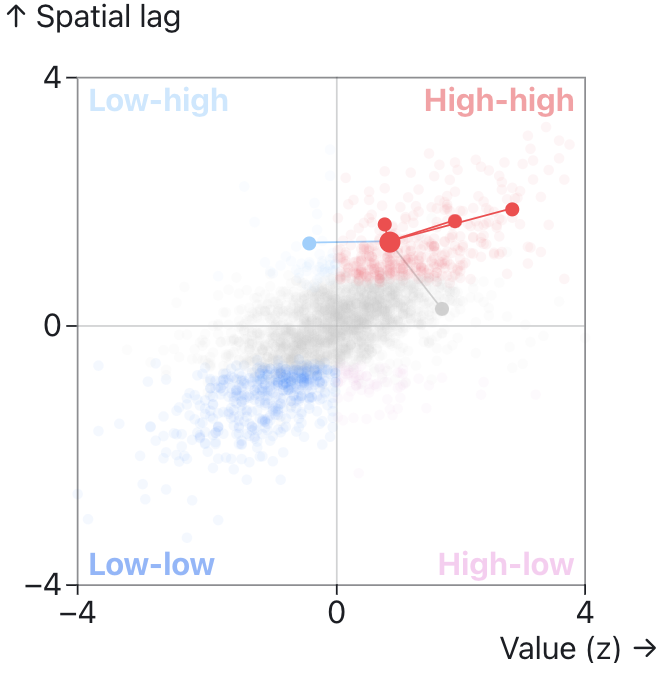}
  \caption{%
    The Moran network scatterplot showing a high-high result connected to three other high-high results, a low-high result, and a non significant result.
  }
  \label{fig:network}
\end{figure}

The Moran network scatterplot (see \cref{fig:network}) enhances the standard Moran scatterplot by visualizing neighbor relationships. As in a standard Moran scatterplot, each point represents a location, plotted by its attribute value (x-axis) and spatial lag (y-axis). However, when the user hovers their mouse over a specific point, lines are drawn connecting it to the points of each of its neighbors' results with line thickness proportional to the weight between those locations. This conveys how a result relates to the results of its spatial neighbors. Line colors indicate neighbor labels, aiding quick inspection. This plot conveys attribute values (E1), spatial lags (E4), and the elements involved in spatial lag calculation: neighboring values (E2) and weight matrix weights (E3).

\begin{figure*}[t]
  \centering 
  \includegraphics[width=.82\textwidth, alt={}]{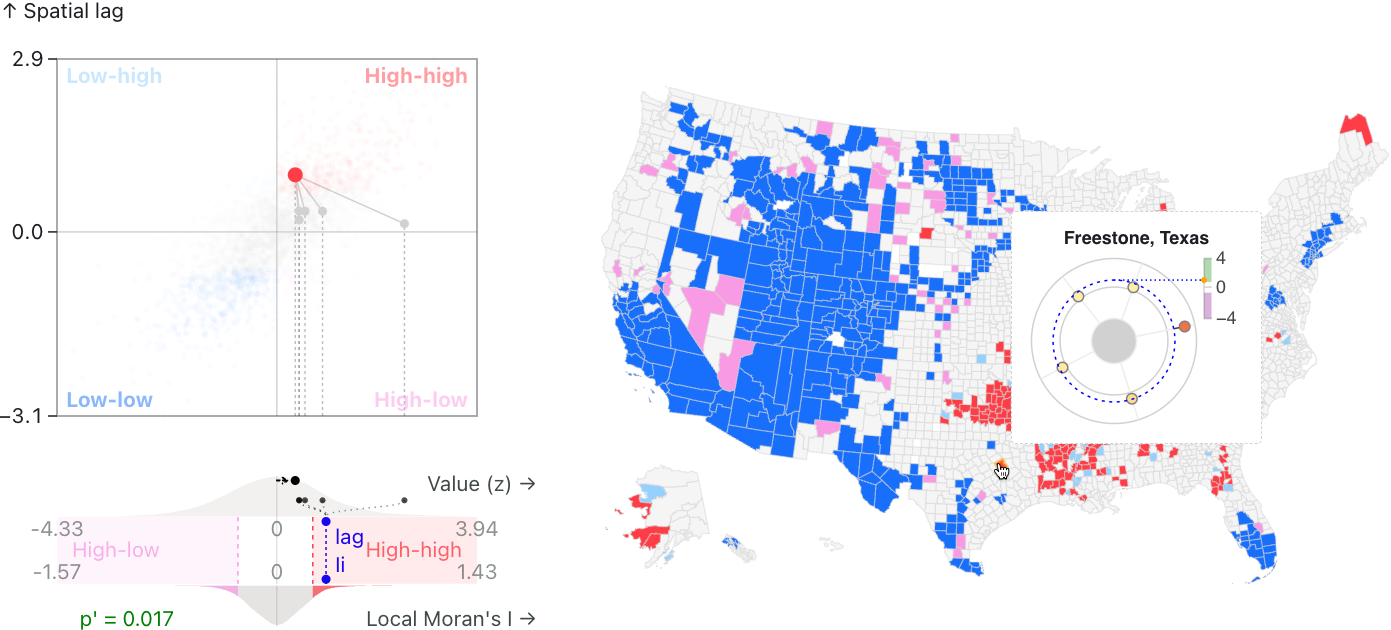}
  \caption{
A screenshot of the MoranPlot dashboard which provides an integrated example of the plots introduced in this work. Here, the user has hovered over the US county of Freestone, Texas and the corresponding result is conveyed in the other plots. The dashboard can be accessed at \url{https://episphere.github.io/moranplot}.
 }
  \label{fig:dash}
\end{figure*}

We extended the Moran's scatterplot in this manner to better contextualize results. When an analyst looks at spatial clustering results, particularly on a cluster choropleth map, they tend to consider groupings of same-label results as a higher-level feature. For instance, the southern tip of Florida in \cref{fig:dash} shows a contiguous group of low-low results --- it is natural to consider these results part of a single cancer mortality cold-spot. However, on a traditional Moran's scatterplot, this grouping would be obscured. We further enhanced this feature by showing the full connection of neighboring results with identical labels when the user double clicks on a point. 

\subsection{Spatial Lag Radial Plot}

\begin{figure}[h]
  \centering 
  \includegraphics[width=0.71\columnwidth, alt={A figure showing the spatial lag radial plot.}]{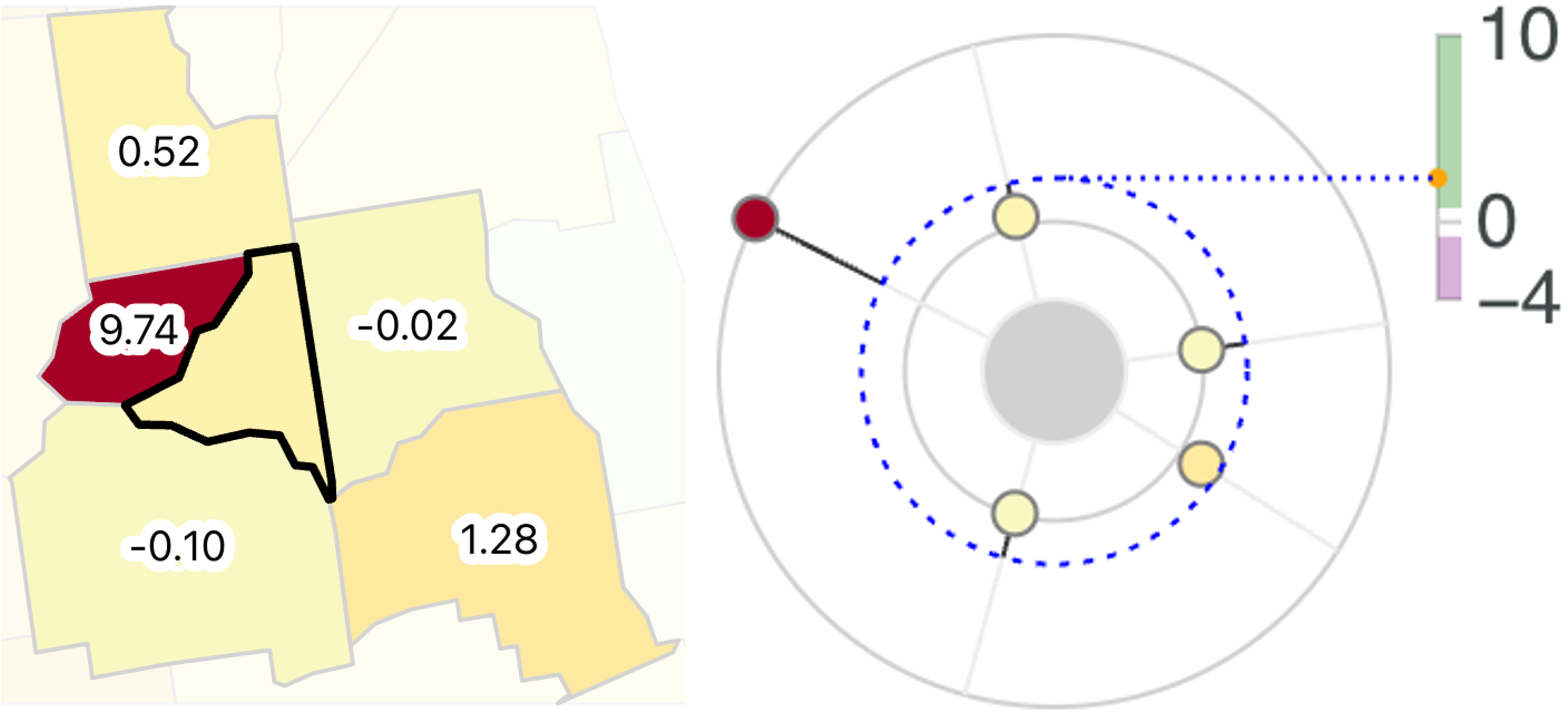}
  \caption{%
    A spatial lag radial plot (right) showing a location with positive spatial autocorrelation. Note how the the substantially high value to the North West is primarily responsible for the lag exceeding the threshold for significance.
  }
  \label{fig:radial}
\end{figure}

The spatial lag radial plot illustrates each neighbor's contribution to the spatial lag (E2, E3, and E4). Neighbors are plotted on a circular axis, their distance from the center proportional to their z-score normalized attribute value. A blue dashed circle represents the spatial lag (E4), its radius is on the same scale as the neighbor points. The neighbor points are connected to this circle by a line to illustrate whether they increase or decrease the lag. A grey hollow circle denotes $z=0$, aiding in comparing the lag to the mean attribute value. A central grey filled circle represents the minimum attribute value across the dataset. Each point's radius is proportional to its weight in the weight matrix, and its angle matches the direction of the corresponding neighbor's centroid from the focal location's centroid.

We designed this plot to represent each neighbor's contribution to the Local Moran's I calculation while maintaining a connection to their location in space. The radial plot supports this goal by aligning an axis for each neighbor to correspond to its direction from the focal location, helping the user to visually connect a neighbor's contribution to the Local Moran's I calculation with its spatial position.

%%%%%%%%%%%%%%%%%%%%%%%%%%%%%%%%%%%%%%%%%%%%%%%%%%%%%%%%%%%%%%%%
%%%%%%%%%%%%%%%%%%%%%%%%%%%%%%%%%%%%%%%%%%%%%%%%%%%%%%%%%%%%%%%%
\section {Linking the plots}

The Moran dual-density and spatial lag radial plots are designed to show a single Local Moran's I result at a time. However, a typical analysis will produce many such results, and therefore it is useful to have a natural means by which to access each one. Here we will suggest a design linking these plots in a way that emphasizes the connections between them and encourages exploration. We have provided an example dashboard which illustrates this design at \url{https://episphere.github.io/moranplot}. In the dashboard, the network scatterplot and LISA cluster map act as a home for user exploration, with the user's  interactions reflected across the other plots. When the user hovers over a location in either the Moran network scatterplot or the LISA cluster map, the Moran dual-density plot is updated to show the details for the corresponding result. If the user hovers over a location in the map, a graphical tooltip appears showing a lag radial plot for that location. The decision to render this element in a tooltip instead of another panel in the dashboard was to minimize the visual distance between the geospatial information (presented in the map) and the radial orientation, simplifying the user's mental connection between the two. 

In a dashboard, it is important to maintain the user's mental map across plots to encourage a comprehensive understanding of what is displayed \cite{bachDashboardDesignPatterns2022}. A crucial element of this is ensuring that the user understands how corresponding elements across different plots relate to one another. In the example dashboard, we achieve this in a few different ways. The first is that we position the Moran network scatterplot directly above the Moran dual-density plot with the former's upper x-axis aligned with the latter's x-axis --- both show the z-score normalized attribute values. This alignment clearly shows the direct relationship between the two plots. When the user hovers over a specific point in the Moran network scatterplot, vertical lines are drawn connecting that point, and each of its neighbors, to the bottom axis. Due to the aligned x-axes, this directly conveys the link between the highlighted points in the Moran network scatterplot and the points featured in the Moran dual-density plot. 

We used established interactive visualization principles to link these plots in a way that does not overwhelm the user's visual processing capacity, including preserving the user's mental map using shared colors, applying link-and-brush interactions to connect plots, and only displaying detailed information on demand. \cite{robertsStateArtCoordinated2007, cuiVisualAnalyticsComprehensive2019a, thomasVisualAnalyticsAgenda2006, bachDashboardDesignPatterns2022, viauConnectedChartsExplicitVisualization2012, interactiveDataVisualizationBuja1991}.

%%%%%%%%%%%%%%%%%%%%%%%%%%%%%%%%%%%%%%%%%%%%%%%%%%%%%%%%%%%%%%%%
%%%%%%%%%%%%%%%%%%%%%%%%%%%%%%%%%%%%%%%%%%%%%%%%%%%%%%%%%%%%%%%%
\section{Use Case: US Cancer Mortality}

We will now outline a brief example analysis that illustrates how these novel plots may be applied to real world data. The dataset is US county-level age-adjusted cancer mortality from the CDC, aggregated from 2011 to 2020.  
A user will typically begin by inspecting the overview of the results from either the Moran network scatterplot or the LISA cluster map. From the latter, they will see several clusters of high and low rates across the country. If the user inspects the Moran dual-density plots at various locations, they will see that the majority of significant results exhibit a relatively uniform distribution of neighbor values, suggesting the label assignments are robust and not primarily driven by outliers. However, some cases like Freestone, Texas (see \cref{fig:dash}) illustrate that significance can be determined by a single outlier, as seen in both the Moran dual-density plot and the lag radial; the latter reveals the outlier to be Anderson, Texas. These observations would be challenging to ascertain from other plots. 
%%%
From the Moran network plot, the user can get a feel for how results cluster together, and how this relates to the attribute and spatial lag values. In this dataset, significant points further up the x-axis and y-axis tend to be linked to other significant points. This shows a general spatial gradation of values where geospatial centers of significant result clusters tend to have high attribute and lag values than locations on the edges. There are some exceptions: Freestone, Texas (see \cref{fig:dash}), for example, has been assigned a high-high label, but it is connected to four non-significant results; it is an isolated hot-spot cluster center.

%%%%%%%%%%%%%%%%%%%%%%%%%%%%%%%%%%%%%%%%%%%%%%%%%%%%%%%%%%%%%%%%
%%%%%%%%%%%%%%%%%%%%%%%%%%%%%%%%%%%%%%%%%%%%%%%%%%%%%%%%%%%%%%%%
\section{Feedback and Discussion}

In this work, we have introduced three novel plots designed to address gaps in the visualization landscape of the Local Moran's I statistic. In designing the plots, we solicited feedback from two experts in spatial analysis/visualization and three domain experts in epidemiology. All individuals expressed interest in the plot designs and provided suggestions for improvement. One of the spatial experts stated that the dual-density plot could be used to determine when spatial datasets are poorly conditioned for Local Moran's I analysis. The other identified the explicit representation of the connections between a focal location and its neighbors as a particularly useful feature of the dual-density and network plots. Initially, the dual-density plot showed the distribution of spatial lags alongside the distribution of attribute values, but one of the spatial experts expressed that they found this confusing, and so we removed it (it remains an option in the library). This expert also stated that the compactness of information in the dual-density plot could potentially impede understanding for non-spatial experts, which suggests that the design may require some refinement or a simplified alternative. However, the non-spatial experts who provided feedback unanimously said that the plot helped them understand the statistic better. One of the spatial experts and one of the domain experts expressed confusion over the lag radial plot, suggesting the design needs further refinement to become intuitive. Initially, the dual-density plot showed the direct outcomes of a Local Moran's I analysis (positive or negative spatial autocorrelation), but when one of the domain experts expressed confusion at this we changed the default colors to instead correspond to the label assignments (e.g. high-high), and included text to make the meaning of each area explicit. The original coloring is included as an option in the library. This informal feedback has suggested that these plots are a potentially valuable contribution to the Local Moran's I visualization landscape, but a more rigorous evaluation is required to better understand their strengths and limitations.

We believe that the Moran dual-density plot stands out as a particularly novel contribution because it is the first Moran plot design to explicitly depict the main elements of a Local Moran's I calculation. The plot quickly conveys useful insights, including how close the statistic or lag value is to a significance boundary, or how a these values are affected by extreme neighbor attribute values. The plot also  represents how the elements of a Local Moran's I calculation relate to the final result, which could help novice analysts better understand the statistic. 
The other two plots convey subsets of the Local Moran's I calculation elements in different ways, with the aim of providing a more holistic picture of the results when viewed together. To facilitate this, we provided a suggestion for how these plots can be linked using shared visual elements and user interaction. 
In addition to making the calculations clearer at a glance to experienced geospatial analysts, we hope that the focus on interpretability will help those new to Local Moran's I analysis better understand the statistic and how it applies to their dataset. 

The plots introduced in this work have been designed with the assumption that each location will have a small number of spatial neighbors, which is typically the case with the standard contiguous neighbor methodology used for areal data. However, some forms of spatial analysis (e.g. on point location data) may involve a large number of neighbors so it is important to consider how well these plots scale. The dual-density plot uses a single axis of small, semi-transparent dots to represent neighbors and thus should be readable with a large number of neighbors, unless many of those neighbors share a similar value. In this case the y-axis could be used as a way to separate the neighbors (e.g. using jitter). The scalability of the radial plot depends on the geospatial orientation of the neighbors: if there are many in a similar direction the plot will become difficult to read. The network plot could potentially become overwhelmed by connecting lines if the number of neighbors is too large. Furthermore, the network plot's full contiguous result highlighting functionality will lead to visual clutter when the number of results in a result cluster is large.

%%%
In summary, we have introduced three novel plots for visualizing the Local Moran's I statistic, addressing gaps in the existing visualization landscape. We have provided a design that links these plots together in a manner that encourages exploration of Local Moran's I results among both experienced and novel geospatial analysts. 

%the aim of this work is to further the landscape of Local Moran's I visualizations by explicitly representing aspects of the results that have been thus far neglected by other visualization designs, empowering exploration among both experienced and novice geospatial analysts.

\section*{Supplemental Materials}

% TODO: Explicitly mention library

The full source code and documentation are available at \url{https://github.com/episphere/moranplot}. 

A dashboard showing the interactive plots in action can be found at \url{https://episphere.github.io/moranplot} 

An Observable notebook providing an overview of the plots and instructions on how to use the moranplot library can be found at \url{https://observablehq.com/@siliconjazz/moranplot-overview}. 

An Observable notebook describing the Moran dual-density plot in more detail can be found at \url{https://observablehq.com/@siliconjazz/visualizing-local-morans-i-using-dual-density-plot}

\acknowledgments{
The authors wish to thank Dr René Westerholt, Dr Levi Wolf, Dr Neal Freedman, Dr Wayne Lawrence, and Dr Meredith Shiels for their valuable feedback on the proposed plot designs. 

Research reported in this publication was supported by the National Cancer Institute of the National Institutes of Health (CAS 10901).
}

\bibliographystyle{abbrv-doi-hyperref-fix}
\bibliography{references}
\end{document}